\documentclass[12pt]{article}
\usepackage{amsmath}
\usepackage{cite}
\usepackage[dvipdfmx]{graphicx}
\begin{document}    
\begin{flushright}
KANAZAWA-19-04\\
June, 2019
\end{flushright}
\vspace*{1cm}

\renewcommand\thefootnote{\fnsymbol{footnote}}
\begin{center} 
  {\Large\bf Low scale leptogenesis in a hybrid model
of the scotogenic type I and III seesaw}
\vspace*{1cm}

{\Large Daijiro Suematsu}\footnote[1]{e-mail:
~suematsu@hep.s.kanazawa-u.ac.jp}
\vspace*{0.5cm}\\

{\it Institute for Theoretical Physics, Kanazawa University, 
Kanazawa 920-1192, Japan}
\end{center}
\vspace*{1.5cm} 

\noindent
{\Large\bf Abstract}\\
The scotogenic type I and type III seesaw models are good candidates to 
explain the existence of neutrino masses and dark matter simultaneously.
However, since triplet fermions have $SU(2)$ gauge interaction, they cannot be 
out of equilibrium before the electroweak symmetry breaking.
Thus, leptogenesis seems to be difficult within a framework of the 
pure type III seesaw model.  Some extension seems to be required to
solve this fault. 
A model extended by introducing a singlet fermion could be such a simple example.  
If the singlet fermion is in the thermal equilibrium even for its extremely small
neutrino Yukawa coupling, leptogenesis could be shown to occur 
successfully for a rather low mass of the singlet fermion. 
The required mass could be lowered to $10^4$~GeV.   
 
\newpage
\setcounter{footnote}{0}
\renewcommand\thefootnote{\alph{footnote}}

\section{Introduction}
Leptogenesis is considered to be the most promising scenario for the 
generation of baryon number asymmetry in the Universe \cite{fy,leptg}.
In this scenario, lepton number asymmetry produced in some way is transformed 
to the baryon number asymmetry through the sphaleron interaction \cite{sph}.
The lepton number asymmetry  is usually considered to be caused through 
the decay of right-handed neutrinos which appear in the seesaw mechanism 
for the neutrino mass generation \cite{seesaw}. 
If the right-handed neutrinos have no interaction except for neutrino 
Yukawa couplings, both their production in the thermal plasma and their decay
are brought about only through this interaction. 
If these couplings are strong, their production
occurs effectively and they can reach equilibrium at an earlier stage.
However, washout of the generated lepton number asymmetry
is also caused by them effectively. 
On the other hand, if these couplings are weak, their production is ineffective
and their equilibrium value is realized at a later stage although 
the washout effect could be suppressed. 
As a result, only a restricted range of the neutrino Yukawa couplings 
is expected to cause the required baryon number asymmetry via the leptogenesis
successfully. This feature requires the mass of 
the right-handed neutrinos in the ordinary seesaw model 
to be more than $10^9$~GeV \cite{di}
under the constraint of neutrino oscillation data as long as resonant 
leptogenesis \cite{resonant} 
is not supposed. We find the similar feature in the scotogenic type I
seesaw model \cite{ks}, which is a well-known model for both neutrino masses 
and dark matter (DM) \cite{scot1,scot1a}.  
In this model, the right-handed neutrinos whose masses are in TeV ranges 
could have a chance to be each candidate for DM and a mother 
field of leptogenesis \cite{ds}. 

The scotogenic type III seesaw model is known 
as another model which can connect the 
neutrino mass generation and the existence of DM at
low energy regions \cite{scot3}. 
It is a simple extension of the standard model (SM)
by an additional inert doublet scalar $\eta$ and $SU(2)$ triplet fermions 
$\Sigma_\alpha~(\alpha=1-n_\Sigma)$ which could play the same 
role as the right-handed 
neutrinos in the scotogenic type I seesaw model.
If odd parity of a $Z_2$ symmetry imposed on the model 
is assigned to these new fields and all other fields are assumed to have its 
even parity, the neutrino masses are forbidden
at a tree level but they are generated through a one-loop diagram.
This model can have also two DM candidates, a neutral component of $\eta$ and 
the lightest neutral one of $\Sigma_\alpha$, whose stability is guaranteed by
the $Z_2$ symmetry.  
In both cases, one might expect that the decay of the lightest or 
next lightest triplet fermion causes the lepton number asymmetry 
depending on which is the DM since
it violates the lepton number. 
However, this decay is difficult to cause a net lepton number asymmetry 
unfortunately since $\Sigma_\alpha$ are considered to have masses 
near the TeV ranges.
Since the triplet fermions $\Sigma_\alpha$ have $SU(2)$ gauge interaction
differently from the right-handed neutrino,
it cannot be out of equilibrium before the electroweak symmetry 
breaking \cite{2019s}.\footnote{Leptogenesis in the type III seesaw model 
has been studied in \cite{tri-l}. It has been shown that the sufficient lepton number 
asymmetry can be generated as long as the mass of the mother 
triplet fermion is larger than $O(10^9)$ GeV. However, it has also discussed
that successful leptogenesis is not so easy for a much lighter triplet fermion.}   
In that case, their decay cannot satisfy the Sakharov conditions for the generation of
the lepton number asymmetry.
In this paper, we try to extend the scotogenic type III seesaw model to
incorporate the leptogenesis in it in a self-contained way assuming that 
the $\Sigma_\alpha$ mass is much smaller than $O(10^9)$ GeV.
In that extension, the sufficient baryon number asymmetry is found to be 
produced by a mother fermion with a mass of $O(10^4)$~GeV. 

The paper is organized as follows. In the next section, we introduce a scotogenic
type III seesaw model and give a brief review of the neutrino mass generation 
and the DM abundance in it. After that, its extension  
is discussed by introducing a singlet fermion and we address how 
it makes leptogenesis possible.
In section 3, the leptogenesis is studied quantitatively to show that
it could occur for a rather low mass mother fermion. 
The paper is summarized in section 4.

\section{A hybrid scotogenic model}
\subsection{Neutrino mass and DM abundance in scotogenic type III seesaw}
The scotogenic type III seesaw model \cite{scot3} is characterized by the neutrino 
Yukawa couplings of $SU(2)$ triplet fermions $\Sigma_\alpha$ with 
a hypercharge $Y=0$ and an inert doublet scalar $\eta$ with $Y=-1$, which are
given as 
\begin{equation}
-{\cal L}_\Sigma=\sum_{\alpha=1}^{n_\Sigma}
\left(\sum_{i=e,\mu,\tau}h_{i\alpha}\bar\ell_{L_i}\Sigma_\alpha\eta +
\frac{1}{2}M_\alpha{\rm tr}(\bar\Sigma_\alpha\Sigma_\alpha^c)
+{\rm h.c.} \right),
\label{yukawa}
\end{equation}
where $\Sigma_\alpha$ is defined by
\begin{equation}
\Sigma_\alpha\equiv\sum_{a=1}^3\frac{\tau^a}{2}\Sigma^a_\alpha=\frac{1}{2}\left(
\begin{array}{cc} \Sigma^0_\alpha& \sqrt 2\Sigma^+_\alpha \\ 
\sqrt 2\Sigma^-_\alpha & -\Sigma^0_\alpha\\
 \end{array}\right).
\end{equation}
The scalar potential of the model is given by
\begin{eqnarray}
V&=&m_\phi^2\phi^\dagger\phi+ m_\eta^2\eta^\dagger\eta+
\lambda_1(\phi^\dagger\phi)^2+\lambda_2(\eta^\dagger\eta)^2+
\lambda_3(\phi^\dagger\phi)(\eta^\dagger\eta)+
\lambda_4(\phi^\dagger\eta)(\eta^\dagger\phi) \nonumber \\
&+& \frac{\lambda_5}{2}\left[(\eta^\dagger\phi)^2+(\phi^\dagger\eta)^2\right],
\label{pot}
\end{eqnarray}
where $\phi$ is an ordinary Higgs doublet scalar.
Since we impose a $Z_2$ symmetry for which only $\Sigma_\alpha$ and $\eta$ 
have odd parity and all other fields are assigned even parity, 
their allowed interaction terms except for gauge interactions 
are restricted to the ones listed in eqs.~(\ref{yukawa})
and (\ref{pot}).  

\input epsf
\begin{figure}[t]
\begin{center}
\epsfxsize=14cm
\leavevmode
\epsfbox{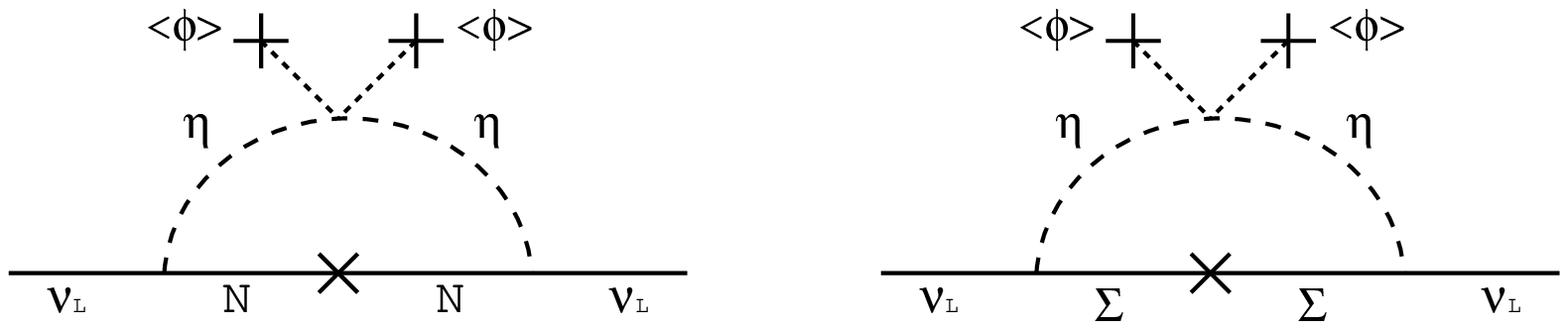}
\end{center}
\vspace*{-13mm}
    {\footnotesize {\bf Fig.~1} (a) A one-loop diagram for the neutrino mass
      generation in the scotogenic type I seesaw.
(b) A one-loop diagram for the neutrino mass generation
      in the scotogenic type III seesaw. }  
\end{figure}

This $Z_2$ symmetry brings about interesting features in the model. 
Since $\eta$ is assumed to have no vacuum expectation value, the $Z_2$ 
symmetry remains as an exact one. Thus, the neutrinos cannot have masses 
at a tree level. However, as the scotogenic type I seesaw model shown in
Fig.~1(a), the neutrino masses are generated by a one-loop diagram shown 
in Fig.~1(b),
in which the right-handed neutrino $N$ in the former is replaced 
by $\Sigma_\alpha$. The mass induced through this diagram is estimated as
\begin{equation}
{\cal M}_{ij}=\sum_{\alpha=1}^{n_\Sigma}
\frac{h_{i\alpha}h_{j\alpha}\lambda_5\langle\phi\rangle^2}
{32\pi^2M_\alpha}\left[\frac{M_\alpha^2}{M_\eta^2-M_\alpha^2}
\left(1+\frac{M_\alpha^2}{M_\eta^2-M_\alpha^2}\right)
\ln\frac{M_\alpha^2}{M_\eta^2}\right],
\label{nmass}
\end{equation}  
where $M_\eta^2=m_\eta^2+(\lambda_3+\lambda_4)\langle\phi\rangle^2$.
If we note that only two triplet fermions are enough to explain the neutrino 
oscillation data, Yukawa coupling constants for the remaining ones can 
be very small so as not to contribute to the neutrino mass generation substantially. 
Taking account of it, we confine our study here to the minimal case $n_\Sigma=2$.
An interesting feature of the model is that both $M_\alpha$ 
and $M_\eta$ can take much smaller values in comparison with typical ones for 
the right-handed neutrino masses in the ordinary type I seesaw model 
as long as $|\lambda_5|\ll 1$ is satisfied.  

Another interesting feature is that the model could explain a required 
value of the DM abundance.
The model has two DM candidates as mentioned above, that is, 
the lightest $\Sigma^0_\alpha$ and the lightest neutral component of $\eta$. 
Both of them have the $Z_2$ odd parity.
In this paper, 
we focus our study on a case where $\eta$ is DM.\footnote{The study 
of a case in which $\Sigma^0$ is DM can be found in \cite{scot3,tridm}.}
This DM candidate has been extensively studied in many articles 
\cite{etadm,ks}. 
There, it has been proved that the lightest neutral component of $\eta$ 
with the mass of $O(1)$~TeV can realize the required DM relic density easily.
In fact, since the co-annihilation among the components of $\eta$ could be 
effective, $\Omega h^2=0.12$ can be obtained for suitable values of 
quartic couplings $\lambda_3$ and $\lambda_4$ without serious fine tuning.
In the following discussion, we just assume $M_\eta=O(1)$~TeV which can 
guarantee the DM abundance.

These are common features to the scotogenic type I seesaw model.
However, a problem is caused in the leptogenesis through a nature of
$\Sigma_\alpha$, which has $SU(2)$ gauge interactions other than 
the neutrino Yukawa couplings given in eq.~(\ref{yukawa}). 
As a result, the $\Sigma_\alpha$ decay cannot generate the lepton number 
asymmetry differently from the right-handed neutrino decay in the 
scotogenic type I seesaw model. This is because they cannot leave the thermal 
equilibrium until a scale of the electroweak symmetry breaking as noted before.
In order to remedy this fault and make the leptogenesis available 
in this framework,
we have to consider some extension of the model.

\subsection{A simple extension of the model}
As a simple extension,\footnote{The hybrid model of the type I and type III is considered in a different context \cite{hyb}.}
we consider to introduce a $Z_2$ odd  
singlet fermion $N$ and add several new terms to the Lagrangian given 
in eq.~(\ref{yukawa}) such that\footnote{Although masses of $\Sigma_\alpha$ 
and $N$ could be supposed to be generated by a vacuum expectation value of 
$S$ \cite{2019s}, they are assumed to be independent parameters for simplicity
in this study.}
\begin{equation}
-{\cal L}_N=\sum_{i=e,\mu,\tau}h_i^N\bar\ell_{L_i}N\eta + 
\frac{1}{2}M_N\bar NN^c+ \frac{1}{2}y_NS\bar NN^c + 
\sum_{\alpha=1}^2\frac{1}{2}y_\alpha{\rm tr}(\bar\Sigma_\alpha\Sigma_\alpha^c)S
+{\rm h.c.},
\label{exlag}
\end{equation}
where $S$ is a $Z_2$ even real scalar which has potential 
$\frac{1}{4}\lambda_SS^4+\frac{1}{2}m_S^2S^2$. 
The mass $m_S$ is assumed to satisfy $m_S \gg M_\alpha>M_N$.
This model can be considered as a hybrid model of the two types of 
scotogenic model since the neutrino masses could be generated through 
the two types of diagram given in Fig.~1. However,
if the coupling constant $h_i^N$ is sufficiently small, the neutrino mass formula (\ref{nmass}) is not affected by this extension.
On the other hand, the smallness of Yukawa coupling $h_i^N$ could make 
the substantial $N$ decay start at a low temperature such as $T\ll M_N$ where 
$\Gamma_N~{^>_\sim}~H$ is
realized for the $N$ decay width $\Gamma_N=\sum_i\frac{h_i^{N 2}}{8\pi}M_N$ 
and the Hubble parameter $H^2=\frac{\frac{\pi^2}{30}g_\ast T^4}{3M_{pl}^2}$.
The decay before reaching this temperature region is out-of equilibrium. 
Thus, as long as $N$ has already been in the thermal equilibrium 
at a high temperature $T>M_N$ through a certain interaction, 
it could generate the lepton number asymmetry efficiently. 

At first, we address how $N$ could be in the thermal 
equilibrium in such a case that its Yukawa couplings $h_i^N$ are very small.
We suppose that $S$ has a non-minimal coupling with Ricci 
scalar such as $\frac{\xi}{2} S^2 R$. In that case, $S$ could play a role of inflaton 
in the same way as Higgs inflation \cite{higgsinf,h-inf}.
This inflation is expected to explain the present observational 
data for the CMB well for appropriate values of $\lambda_S$ and $\xi$. 
Since its detail is not crucial for the present purpose, we confine the present
discussion to the estimation of reheating temperature 
only.\footnote{Several inflation scenarios have been discussed in the scotogenic 
type I seesaw model extended by a singlet scalar \cite{scot-inf}. }
The reheating is dominantly caused 
by the $S$ decay to $\Sigma_{1.2}$ pairs through 
the couplings in eq.~(\ref{exlag}) in the case $y_{1,2} > y_N$. 
There, the reheating temperature can be estimated from $H\simeq \Gamma_S^D$ 
by using both the Hubble parameter $H$ and the decay width $\Gamma^D_S$ 
of $S$ such as
\begin{equation}
T_R\simeq 4\times 10^{11}\left(\frac{y_\Sigma}{10^{-2}}\right)
\left(\frac{m_S}{10^{10}~{\rm GeV}}\right)^{1/2}~{\rm GeV},
\label{reheat}
\end{equation}  
where $y_{1,2}=y_\Sigma$ is assumed and $g_\ast=121.5$ is used for relativistic 
degrees of freedom in the model.
Here, it is important to note that $N$ is pair-produced in the thermal plasma 
through the scattering of $\Sigma_{1,2}$ pairs mediated by $S$
even if the Yukawa coupling constants $h_i^N$ are sufficiently small.
In that case, the Yukawa coupling constants $h_i^N$ could be irrelevant 
to the determination of the abundance of $N$. 
This is a completely different situation from the pure scotogenic type I 
seesaw case \cite{ks}.
On the other hand, using the assumption $\frac{m_S}{2} > M_{1,2}>M_N$, 
we can roughly estimate the freeze-out temperature of this scattering 
process from $H\simeq \Gamma^S_{\Sigma_\alpha\Sigma_\alpha\rightarrow NN}$,
where $\Gamma^S_{\Sigma_\alpha\Sigma_\alpha\rightarrow NN}$ is
the reaction rate for $\Sigma_\alpha\Sigma_\alpha\rightarrow NN$, such as
\begin{equation}
T_D\simeq 2\times 10^{10}\left(\frac{10^{-2}}{y_\Sigma}\right)^{2/3}
\left(\frac{10^{-2}}{y_N}\right)^{2/3}
\left(\frac{m_S}{10^{10}~{\rm GeV}}\right)^{4/3}~{\rm GeV}.
\label{freeze}
\end{equation}
Eqs.~(\ref{reheat}) and (\ref{freeze}) suggest that $N$ could reach the 
thermal equilibrium at a certain temperature T such as $T_D<T<T_R$.
After that, it decouples from the thermal plasma at $T<T_D$ and 
starts the out-of-equilibrium decay to $\ell\eta^\dagger$ to generate 
the lepton number asymmetry.

\begin{figure}[t]
\begin{center}
\epsfxsize=7.5cm
\leavevmode
\epsfbox{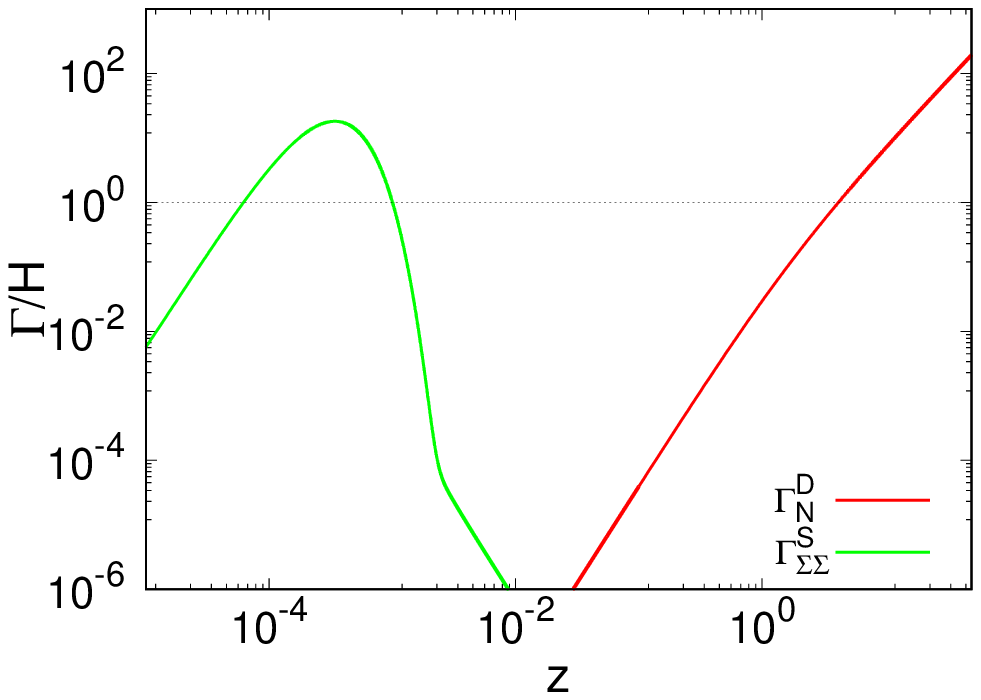}
\hspace*{5mm}
\epsfxsize=7.5cm
\leavevmode
\epsfbox{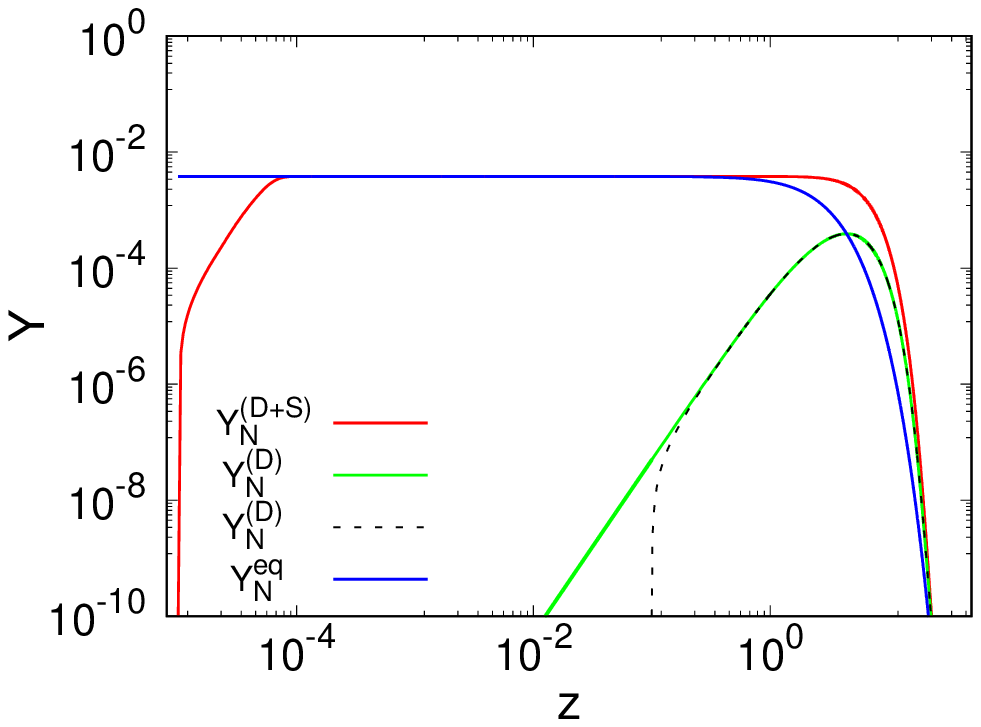}
\end{center}
\vspace*{-3mm}
{\footnotesize {\bf Fig.~2}~~The left panel shows 
the reaction rate $\Gamma$ normalized by the Hubble parameter $H$ 
which is relevant to the $N$ production, that is,
the 2-2 scattering $\Sigma_\alpha\Sigma_\alpha \rightarrow NN$~
($\Gamma^S_{\Sigma\Sigma}$) and
the decay $N \rightarrow \ell\eta^\dagger$~($\Gamma_N^D$).
The right panel shows the evolution of $Y_{N}$ for an initial condition 
$Y_N(z_i)=0$ with $z_i=z_R\left(\equiv\frac{M_N}{T_R}\right)$. 
$Y_N^{\rm eq}$ represents the thermal equilibrium value.
Both $\gamma_D$ and $\gamma_{\Sigma\Sigma}$ are 
taken into account  in $Y_N^{(D+S)}$ but only $\gamma_D$ is 
taken into account in $Y_N^{(D)}$. 
As a reference, $Y_N^{(D)}$ is plotted for the case $z_i=0.1$ using a black
dashed line. }
 \end{figure}

In order to confirm that this scenario works, as an example, we fix the relevant 
parameters as follows, 
\begin{eqnarray}
&&y_\Sigma=10^{-1.5}, \quad y_N=10^{-2}, \quad h_i^N=10^{-6}, 
\quad  m_S=10^{10}~{\rm GeV},
\nonumber \\
&&M_1=10^7~{\rm GeV}, \quad M_2=10^8~{\rm GeV}, 
\quad M_N=10^6~{\rm GeV}.
\label{para1}
\end{eqnarray}
For these parameters, the reheating temperature obtained through the $S$ 
decay can be estimated as $T_R\simeq 4\times 10^{11}$~GeV from 
eq.~(\ref{reheat}). 
To examine the evolution of the number density of $N$, 
we solve the Boltzmann equation for the number density of $N$ 
\cite{kt},
\begin{equation}
\frac{dY_{N}}{dz}=-\frac{z}{sH(M_N)}
\left(\frac{Y_{N}}{Y_{N}^{\rm eq}}-1\right)\left\{\gamma_D
+\left(\frac{Y_{N}}{Y_{N}^{\rm eq}}+1\right)\gamma_{\Sigma\Sigma}\right\},
\label{beqn}
\end{equation}
where $\Sigma_{1,2}$ are supposed to be in the thermal equilibrium.
$z$ and $H(M_N)$ are defined as 
$z\equiv\frac{M_N}{T}$ and 
$H(M_N)\equiv 0.33g_\ast^{1/2}\frac{M_N^2}{M_{\rm pl}}$. 
$Y_N$ is defined as $Y_N\equiv\frac{n_N}{s}$ by using  
the $N$ number density $n_N$ and the entropy density $s$.
$Y_N^{\rm eq}$ represents its equilibrium value.
$\gamma_D$ and $\gamma_{\Sigma\Sigma}$ stand for the reaction density
of the $N$ decay and the 2-2 scattering 
$\Sigma_\alpha\Sigma_\alpha \rightarrow NN$, respectively \cite{reactiond}. 

The solution of eq.~(\ref{beqn})  is plotted in Fig.~2. 
In the left panel, the ratio of each reaction rate to the Hubble parameter 
$\frac{\Gamma}{H}$ is plotted as a function of $z$.
They are relevant to the production of $N$.
In the right panel, the evolution of $Y_N$ is plotted as a function of $z$ 
for two cases, that is, in $Y_N^{(D+S)}$ where both the inverse decay of $N$ and
the 2-2 scattering of a $\Sigma_\alpha$ pair are taken into account, 
but in $Y_N^{(D)}$ where the former is taken into account alone.
The comparison of both panels suggests that the thermal equilibrium abundance 
of $N$ is realized when the 2-2 scattering reaches the equilibrium, 
and it is kept still after the 2-2 scattering leaves the equilibrium.
The out-of-equilibrium decay of $N$ starts at $z>1$.
Since $Y_N(z_R)=0$ is assumed as an initial value at $T_R$,
the right panel shows that $N$ is efficiently produced by the 2-2 
scattering and $Y_N^{(D+S)}$ reaches the equilibrium value $Y_N^{\rm eq}$ 
at a higher temperature compared with no scattering one $Y_N^{(D)}$.
We can also find from this panel that the out-of-equilibrium decay of $N$ 
could start at a larger $Y_N^{\rm eq}$ value in the $Y_N^{(D+S)}$ case 
than the one in the $Y_N^{(D)}$ case. The difference is found to be 
one order of magnitude in this example. 
This feature does not depend on $T_R$ as long as $Y_N$ reaches 
its equilibrium value before $z\sim 1$. 
It suggests that $N$ could be a good mother fermion for 
the lepton number asymmetry in this extended model.
Since both the mass and the couplings of $N$ are free from the neutrino 
mass constraint, a window might be opened for the low scale leptogenesis.  

In the same panel, as a reference, $Y_N$ is plotted also for the case 
$Y_N^{(D)}(0.1)=0$ by a dashed line.
It shows that $Y_N^{(D)}$ immediately reaches the same 
value for the case $Y_N^{(D)}(z_R)=0$. 
This suggests that we can take a much larger $z$ than $z_R$ as a starting
point for the analysis of the Boltzmann equation. 
Taking account of it, we discuss
a possibility of the low scale leptogenesis caused by the $N$ decay 
quantitatively in the next section.

\section{Leptogenesis}
The $N$ decay could satisfy the Sakharov condition and then 
it generates the lepton number asymmetry, which is converted to the
baryon number asymmetry through the sphaleron process. 
If the sphaleron is in the thermal equilibrium, the baryon number $B$ is
found to be related with $B-L$ as $B=\frac{8}{23}(B-L)$ in the present model
by using the chemical equilibrium condition  \cite{chem}.
If we use this relation for $Y_B$ and $Y_{B-L}$ which 
are defined as $Y_B\equiv\frac{n_B}{s}$ and $Y_{B-L}\equiv\frac{n_B-n_L}{s}$ 
by using the entropy density $s$,
$Y_B$ in the present Universe is found to be obtained from $Y_{B-L}$ which is 
produced through the $N$ decay as
\begin{equation}
Y_B=\frac{8}{23}Y_{B-L}(z_{EW}),
\end{equation}
where the dimensionless parameter $z$ is defined in the previous part 
and $z_{EW}$ is fixed by the sphaleron decoupling temperature $T_{EW}$ as
$z_{EW}=\frac{M_N}{T_{EW}}$. 

\begin{figure}[t]
\begin{center}
\epsfxsize=12cm
\leavevmode
\epsfbox{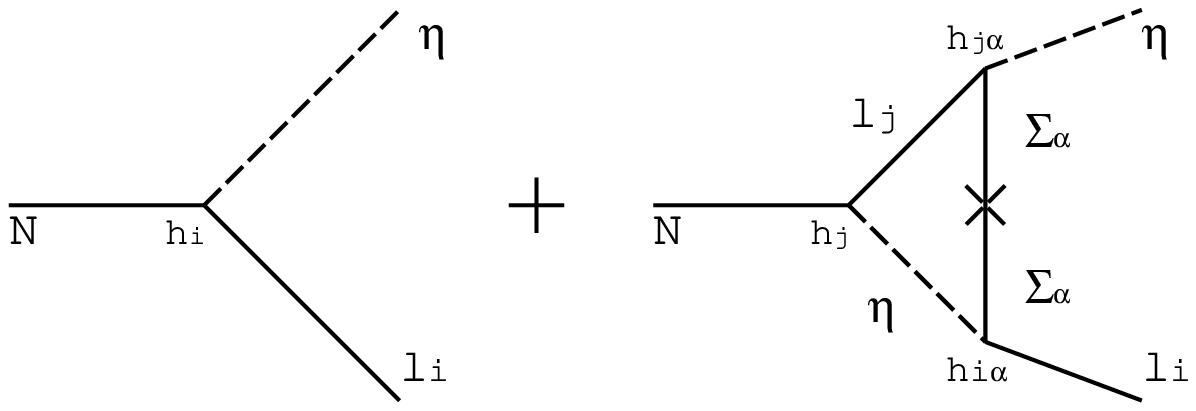}
\end{center}
\vspace*{-5mm}
    {\footnotesize {\bf Fig.~3} ~Diagrams of the lepton number violating 
$N$ decay.  The $CP$ asymmetry is induced by the interference 
between a tree and a one-loop diagram. }  
\end{figure}

The $CP$ asymmetry in the $N$ decay is dominantly caused by the 
interference between a tree diagram and a one-loop vertex diagram
which has $\Sigma_\alpha$ in an internal line. They are shown in Fig.~3.
It is calculated as \cite{epsilon} 
\begin{eqnarray}
\varepsilon&\equiv&\frac{\Gamma(N\rightarrow\ell\eta^\dagger)-
\Gamma(N^c\rightarrow\bar\ell\eta)}{\Gamma(N\rightarrow\ell\eta^\dagger)+
\Gamma(N^c\rightarrow\bar\ell\eta)} \nonumber \\
&=&\frac{3}{64\pi[\frac{3}{4}+\frac{1}{4}(1-\frac{M_\eta^2}{M_N^2})^2]}
\sum_{\alpha=1,2}
\frac{{\rm Im}\left[\left(\sum_{i=e,\mu,\tau}h_i^Nh^\ast_{i\alpha}\right)^2\right]}
{\sum_{i=e,\mu,\tau}h_i^Nh_i^{N\ast}}
G\left(\frac{M_\alpha^2}{M_N^2}, \frac{M_\eta^2}{M_N^2}\right),
\label{cp}
\end{eqnarray}
where $G(x,y)$ is defined as 
\begin{eqnarray}
G(x,y)&=&\frac{5}{4}F(x,0)+\frac{1}{4}F(x,y)+\frac{1}{4}(1-y)^2
\left[F(x,0)+F(x,y)\right], \nonumber \\
F(x,y)&=&\sqrt{x}\left[1-y-(1+x)\ln\left(\frac{1-y+x}{x}\right)\right].
\end{eqnarray}
In order to estimate the lepton number asymmetry quantitatively, 
we need to fix a flavor structure 
of neutrino Yukawa coupling constants $h_{\alpha i}$ and 
$h^N_i$.\footnote{As stressed in \cite{kai}, $\varepsilon$ does not depend on
the PMNS matrix. However, the PMNS matrix could affect  the reaction rate of the
processes which contribute to the washout of the generated lepton number
asymmetry.} Here, we adopt the tri-bimaximal flavor structure as an example. 
In the previous works \cite{ks}, we find that 
tri-bimaximal flavor structure does not cause serious effect 
in the study of leptogenesis compared with 
the one where non-zero $\theta_{13}$ is taken into account.\footnote{
Although the model is different from the one studied in \cite{ks},
the neutrino mass generation is the same except that 
$N$ is replaced by $\Sigma^0$ as shown in Fig.~1.  Since non-zero 
$\theta_{13}$ effects on the neutrino Yukawa couplings are 
considered to appear in both models in the same way, 
we can follow the results there. }
We assume \cite{tribi}
\begin{equation}
h_{e1}=0, ~h_{\nu 1}=h_{\tau 1}\equiv h_1; \quad 
h_{e2}=h_{\mu 2}=-h_{\tau 2}\equiv h_2;
\quad
h_e^N=0,~h_\mu^N=h_\tau^N\equiv h_N, 
\end{equation}
where $h_i^N$ is taken to be very small so that it is 
irrelevant to the neutrino mass and mixing.  
By using this flavor structure of the neutrino Yukawa couplings,  
$\varepsilon$ is found to be expressed as
\begin{equation}
\varepsilon=\frac{3|h_1|^2}{32\pi} 
G\left(\frac{M_1^2}{M_N^2}, \frac{M_\eta^2}{M_N^2}\right)\sin(2\varphi_1),
\end{equation}
where $\varphi_1={\rm arg}(h_N)-{\rm arg}(h_1)$.
Here, we should note that $\varepsilon$ could take a larger value compared with
the one in the pure scotogenic type I seesaw model since a singlet fermion
is replaced by a triplet fermion $\Sigma_\alpha$ in an internal line of 
the one-loop diagram.

If $N$ is in the thermal equilibrium, the substantial generation of the lepton number 
asymmetry is expected to start at $z\sim 1$,
where $N$ leaves equilibrium as found in the right panel of Fig.~2.
Thus, $Y_{B-L}(z_{EW})$ might be roughly estimated as 
$Y_{B-L}(z_{EW})\simeq \varepsilon \kappa Y_N^{\rm eq}(1)$
by using the equilibrium expression 
$Y_{N}^{\rm eq}(z)=\frac{45}{2\pi^4g_\ast}z^2K_2(z)$ where $g_\ast$ is the
number of relativistic degrees of freedom at this period and $K_2(z)$ is the
modified Bessel function of the second kind. $\kappa$ stands for the washout 
efficiency for the generated lepton number asymmetry.
Since the present value of $Y_B$ \cite{pdg}
requires $2,4\times 10^{-10}<|Y_{B-L}(z_{EW})|<2.7\times 10^{-10}$,
we find that $\varepsilon$ has to satisfy 
$|\varepsilon|~{^>_\sim}~ 8\times 10^{-8}\kappa^{-1}$ 
from this rough estimation. 
In the case of  $M_1> M_N$, this $\varepsilon$ value requires
\begin{equation}
|h_1| > 8.5\times 10^{-4}\left(\frac{M_1}{10^7~{\rm GeV}}\right)^{1/2}
\left(\frac{10^6~{\rm GeV}}{M_N}\right)^{1/2},
\label{h1}
\end{equation}
if $\varphi_1=\frac{\pi}{4}$ is assumed.
If the out-of-equilibrium decay of $N$ 
starts at $z\sim 1$, its decay width $\Gamma^D_N$ should satisfy 
$H> \Gamma^D_N$ there.  
On the other hand, its decay should be completed at some $z_T$ before reaching  
the sphaleron decoupling temperature $T_{EW}\sim 100$~GeV
and then $H|_{z=z_{EW}}<\Gamma_N^D$ should be satisfied.
In such a case, $Y_N$ can take its equilibrium value at $z>z_T$.
These impose the condition such as
\begin{equation}
6.2\times 10^{-10}\left(\frac{10^6~{\rm GeV}}{M_N}\right)^{1/2} <
h_N < 6.2\times 10^{-6}\left(\frac{M_N}{10^6~{\rm GeV}}\right)^{1/2}.
\label{hn}
\end{equation} 
This suggests that a favored range of $h_N$ becomes narrower for a smaller
value of $M_N$.  

The situation is completely different from the case discussed above,
if $N$ has to be produced only through 
the neutrino Yukawa coupling $h_N$ from an initial value $Y_N(z_R)=0$.
Since the inverse decay rate of $N$ which is a dominant process of 
the $N$ production is proportional to $h_N^2M_N$,  
$Y_N \ge Y_N^{\rm eq}$ can be realized at a much lower temperature such as 
$z > 1$ for small values of $h_N$.
It is found in the right panel of Fig. 2. 
Thus, the substantial lepton number generation starts at a larger 
$z$ where $Y_N(z)$ is much smaller than $Y_N^{\rm eq}(1)$.
This is one of the reasons why the low scale leptogenesis is not so 
easy in the ordinary seesaw model. 
The present model could escape this difficulty since the Yukawa
coupling $h_N$ is irrelevant to both the $N$ production and
the neutrino mass generation.  

In the above discussion, the washout efficiency $\kappa$ for the generated 
lepton number asymmetry is not taken into account quantitatively. 
The lepton number asymmetry could be washed out mainly by the 
lepton number violating 2-2 scattering such as 
$\eta\eta\rightarrow\ell_i\ell_j$ and $\eta\ell_i\rightarrow\eta^\dagger\bar\ell_j$ 
which are mediated by $\Sigma_{\alpha}$ and also the inverse decay of 
$\Sigma_\alpha$ and $N$. 
Since these processes could be heavily suppressed by the Boltzmann factor
at a low temperature region $z\gg 1~(M_1 \gg T)$, we can take $\kappa\simeq 1$ 
if the lepton number asymmetry
is mainly generated at this region. Such a situation is expected to occur 
in a tiny $h_N$ case.
On the other hand, if the lepton number asymmetry is generated at a smaller
$z$ region such as
$z~{^<_\sim}~10$, $\kappa$ could take a smaller value $(\kappa \ll 1)$ there. 
The above washout processes are proportional to $h_{1,2}^4$, $h_{1,2}^2$ 
and $h_N^2$ respectively, while the $CP$ asymmetry $\varepsilon_{1,2}$ 
is proportional to $h_{1,2}^2$.
Thus, we find that the values of $h_{1,2}$ contained in a restricted region 
is favored for the generation of the lepton number asymmetry. 
Such values of $h_{1,2}$ can be realized for a certain 
range of $|\lambda_5|$ as found from the neutrino mass formula (\ref{nmass})
if masses of $\eta$ and $\Sigma_\alpha$ are fixed.

To examine a possibility of the low scale leptogenesis suggested above 
and to estimate the produced baryon number 
asymmetry quantitatively, we solve the Boltzmann equation 
for $Y_L\equiv Y_\ell-Y_{\bar\ell}$.  In the present model, we can use
the equilibrium value $Y_N^{\rm eq}$ as the initial value of $Y_N$. 
It can be realized through the 2-2 scattering of the $\Sigma_\alpha$ pair 
as addressed in the previous part.
The Boltzmann equation analyzed here is\footnote{Since the lepton number 
violation due to the sphaleron is not introduced in this equation, this $Y_L$ 
should be understood as $-Y_{B-L}$.} 
\begin{equation}
\frac{dY_L}{dz}=\frac{z}{sH(M_N)}\left[
\varepsilon\left(\frac{Y_{N}}{Y_{N}^{\rm eq}}-1\right)\gamma_N^D
-\frac{2Y_L}{Y_\ell^{\rm eq}}\left\{\sum_{f=N, \Sigma_\alpha}
\frac{\gamma_f^D}{4} +\gamma_{\eta\ell}
+\gamma_{\eta\eta}\right\}\right], 
\label{bqn}
\end{equation}
where $Y_\ell^{\rm eq}$ stands for the equilibrium value of leptons 
which is expressed as $Y_\ell^{\rm eq}=\frac{45}{\pi^4g_\ast}$. 
$\gamma^D_f$ stands for a reaction density for the decay of the fermion $f$,
and $\gamma_{\eta\ell}$ and $\gamma_{\eta\eta}$ represent the
reaction density for $\eta\ell_i\rightarrow\eta^\dagger\bar\ell_j$ 
and $\eta\eta\rightarrow\ell_i\ell_j$, respectively.

\begin{figure}[t]
\begin{center}
\epsfxsize=7.5cm
\leavevmode
\epsfbox{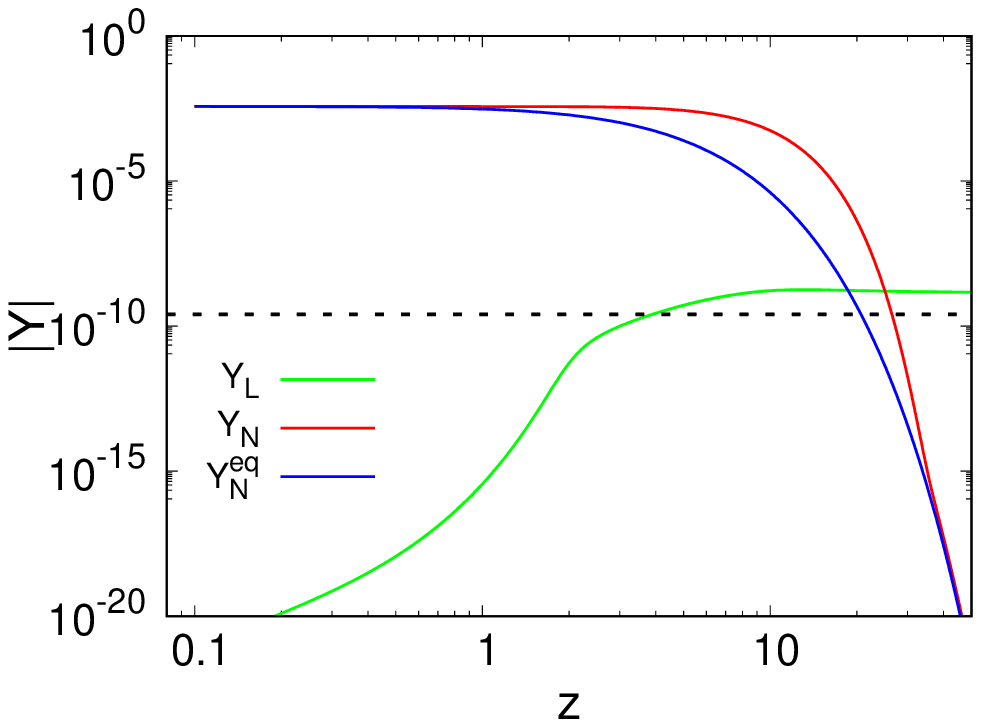}
\hspace{5mm}
\epsfxsize=7.5cm
\leavevmode
\epsfbox{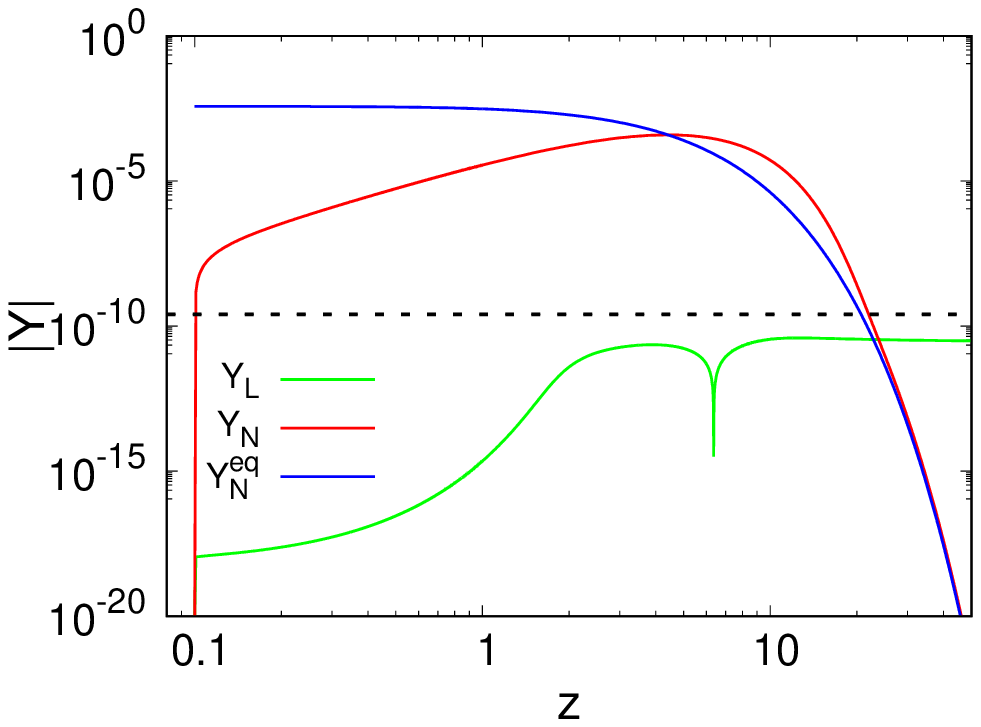}
\end{center}
\vspace*{-3mm}
{\footnotesize {\bf Fig.~4}~~The evolution of $Y_L$ for both initial conditions $Y_N(z_i)=Y_N^{\rm eq}(z_i)$ (left panel) and $Y_N(z_i)=0$ (right panel).
In both panels, the parameters given in (\ref{para1}) are used.
Although we use $z_i=10^{-1}$ in this analysis, the result is not
affected even if $z_i$ is taken to be a smaller value, which has been 
remarked on Fig.~2.  
The black dotted lines in each panel represent the required value of $|Y_L|$.}
 \end{figure}

In order to find the behavior of the generated lepton number asymmetry,
we use the values listed in eq.~(\ref{para1}) for $h_N$, $M_N$, $M_1$ 
and $M_2$.
If we fix $M_\eta$ and $\lambda_5$, the neutrino Yukawa couplings $h_{1,2}$ are
determined through eq.~(\ref{nmass}) by imposing the neutrino oscillation data.
As an example, we fix them at $M_\eta=10^3$~GeV and 
$|\lambda_5|=6\times 10^{-4}$.\footnote{We consider the $\eta$ DM here.
In that case, we have to note that $|\lambda_5|$ is restricted 
by the direct DM search experiments as $|\lambda_5|>5\times 10^{-6}$ \cite{ks,l5}. }
These parameters give the $CP$ asymmetry $|\varepsilon|\simeq 10^{-7}$
for the maximal $CP$ phase.
In Fig.~4, the solutions $Y_N$ and $|Y_L|$ of eqs.~(\ref{beqn}) and 
(\ref{bqn}) are plotted for both initial values
$Y_N(10^{-1})=Y_N^{\rm eq}(10^{-1})$ and $Y_N(10^{-1})=0$.
The left panel shows that a sufficient value of $|Y_L|$
for the explanation of the baryon number asymmetry in the Universe can be 
generated in the former initial value. 
On the other hand, in the latter case plotted in the right panel, 
the generated $|Y_L|$ is found not to reach 
the required value. 
This result can be easily understood by comparing both panels in Fig.~4,
which shows that $Y_N$ in the latter case reaches and leaves the 
equilibrium value at a lower temperature $(z_e\sim 4)$ compared with 
the former case $(z_e\sim 1)$. It directly results in a smaller value of $|Y_L|$
since it can be approximately estimated from 
$Y_L\simeq \varepsilon\kappa Y_N^{\rm eq}(z_e)$ with the same $\kappa$. 
This example suggests that the leptogenesis could occur successfully 
for a rather small mass of the mother fermion in the present model.
At a smaller $h_N$ region, especially, the sufficient baryon number asymmetry 
is expected to be obtained, since the sufficiently late decay of $N$ 
allows almost all the generated lepton number asymmetry to
escape the washout $(\kappa\simeq 1)$ and  be preserved.

\begin{figure}[t]
\begin{center}
\epsfxsize=7.5cm
\leavevmode
\epsfbox{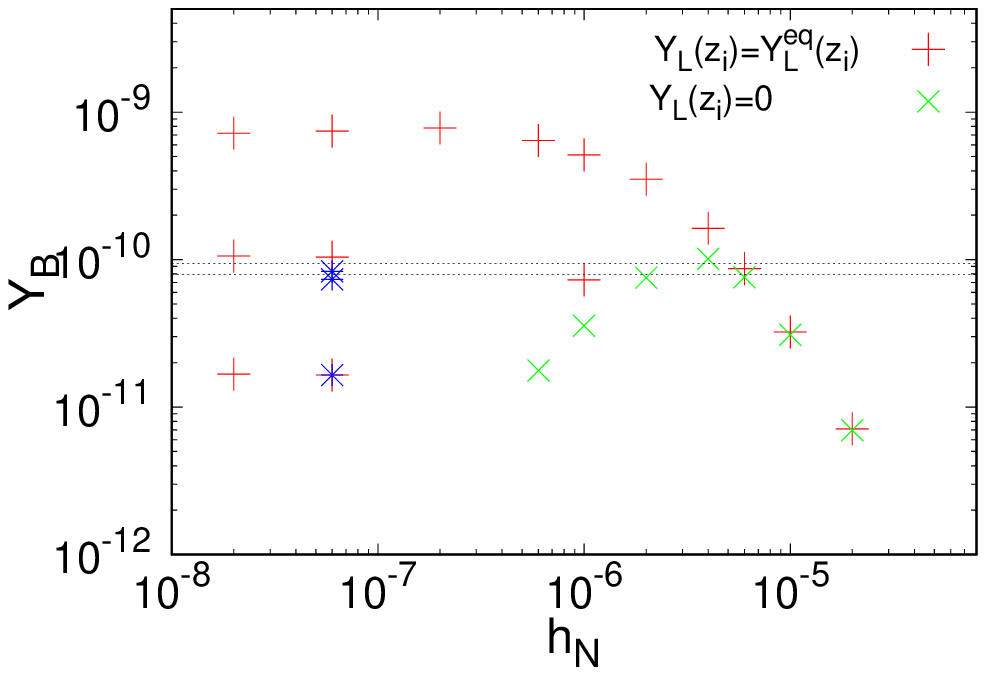}
\hspace*{5mm}
\epsfxsize=7.5cm
\leavevmode
\epsfbox{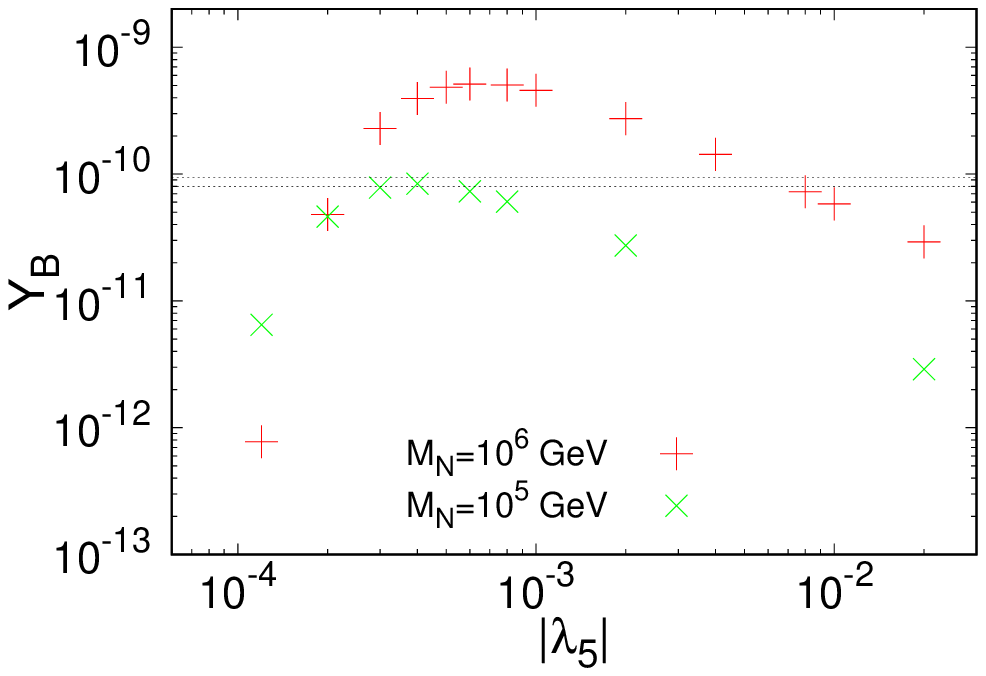}
\end{center}
\vspace*{-3mm}
{\footnotesize {\bf Fig.~5}~~The dependence of $Y_B(z_{EW})$ on $h_N$ and 
$|\lambda_5|$.  In both panels, $z_i=0.1$ is taken and $M_{1,2}$ is assumed to be 
$M_\alpha=10^\alpha M_N$ except for some cases.
The left panel shows the $h_N$ dependence of $Y_B(z_{EW})$ 
in both initial values of $Y_N$.
$M_N$ and $|\lambda_5|$ are fixed at $M_N=10^6$~GeV and 
$|\lambda_5|=6\times 10^{-4}$. 
At $h_N=10^{-6}, ~6\times 10^{-8}$ and $2\times 10^{-8}$, we plot $Y_B(z_{EW})$ 
also for $M_N=10^5$ and $10^4$ GeV downward.
Asterisks at $h_N=6\times 10^{-8}$ display $Y_B(z_{EW})$ for 
$M_N=10^4$ GeV, $M_1=10^5$ GeV and $M_2=2\times 10^5$ GeV by
changing $|\lambda_5|$ as $6\times 10^{-3}, 10^{-4}$ and 
$6\times 10^{-4}$ downward.
The right panel shows the $|\lambda_5|$ dependence of $Y_B(z_{EW})$ 
at $h_N=10^{-6}$. The initial condition is fixed at $Y_N(z_i)=Y_N^{\rm eq}(z_i)$.}
 \end{figure}

A crucial feature of the leptogenesis is controlled by the coupling constants 
$h_N$ and $\lambda_5$ in this model.
In order to clarify it, in Fig.~5 we show the dependence of $Y_B(z_{EW})$ on 
these parameters by fixing the remaining parameters to some typical values. 
In the left panel, $Y_B(z_{EW})$ is plotted for various values of $h_N$ by using 
both initial values $Y_N(10^{-1})=Y_N^{\rm eq}(10^{-1})$ and $Y_N(10^{-1})=0$.
As discussed in eq~(\ref{hn}), the coupling $h_N$ determines a period 
where $N$ is in the out-of-equilibrium state. 
Taking account of this, the $Y_B(z_{EW})$ behavior shown in this panel 
can be explained as follows.
In the case $Y_N(10^{-1})=Y_N^{\rm eq}(10^{-1})$,
the smaller $h_N$ makes the substantial $N$ decay be delayed 
until a low 
temperature where the washout processes are frozen out. 
As a result,  almost all the lepton 
number asymmetry generated through the $N$ decay is transformed to $Y_B$
independently of the $h_N$ value.  
It explains the almost constant behavior of $Y_B(z_{EW})$ at a small $h_N$ region
such as $h_N~{^<_\sim}~2\times10^{-7}$.
In the region $h_N~{^<_\sim}~4\times 10^{-6}$, 
the sufficient $Y_B$ can be obtained for the $Y_N(10^{-1})=Y_N^{\rm eq}(10^{-1})$ 
case, since the washout due to the inverse decay of $N$ is suppressed. 
On the other hand, in the $Y_N(10^{-1})=0$ case, $Y_N>Y_N^{\rm eq}$ tends to be realized at a later period such as $z\gg 1$. 
For a such region of $z$, $Y_N^{\rm eq}$ is too small 
to generate the sufficient $Y_B(z_{EW})$. 
In the region $h_N~{^>_\sim}~ 5\times 10^{-6}$, 
$Y_N>Y_N^{\rm eq}$ starts at $z\sim 1$ commonly for both 
initial values, and then the same $Y_B(z_{EW})$ value is obtained for them.
Although $Y_N$ could be sufficiently large in this case, the washout due to the 
inverse decay of $N$ is effective for this range of $h_N$ 
and then $Y_B(z_{EW})$ is difficult 
to reach a required value. 
Here, it may be useful to note that the required $Y_B(z_{EW})$ could be obtained 
for a suitable value of $h_N$ even in a situation 
$Y_N(10^{-1})=0$ and $M_N<10^8$~GeV,
as found in the left panel.     
It is considered to be caused by a hybrid nature of the model which
makes the $CP$ asymmetry $\varepsilon$ larger compared with 
the pure scotogenic type I seesaw model \cite{ks}.

In the right panel of Fig.~5, $Y_B(z_{EW})$ is plotted for various values of $|\lambda_5|$ 
for two values of $M_N$. 
In this calculation, we choose $h_N=10^{-6}$ and then 
the washout is considered to be 
mainly caused by $\Sigma_\alpha$. 
The figure shows that the $|\lambda_5|$ values included in a restricted 
region can generate a sufficient amount of $Y_B(z_{EW})$.
The coupling $\lambda_5$ determines both magnitudes of the 
$CP$ asymmetry $\varepsilon$ and the washout efficiency $\kappa$ through the
neutrino Yukawa couplings $h_{1,2}$. 
A larger $|\lambda_5|$ gives the smaller $h_{1,2}$ under the constraint of the 
neutrino oscillation data. 
It explains the $Y_B(z_{EW})$ behavior presented in this figure.

\begin{figure}[t]
\begin{center}
\footnotesize
\begin{tabular}{ccc|cccc}
$M_1$(GeV) & $M_2$(GeV) & $|\lambda_5|$ & $h_1$ & $h_2$ & 
$|\varepsilon|$ & $Y_B(z_{EW})$ \\ \hline
$2\times 10^4$& $4\times 10^4$ &  $6\times 10^{-5}$ &$4.2\times 10^{-3}$&
$1.8\times 10^{-3}$& $2.4\times 10^{-7}$ &$4.8\times 10^{-11}$\\
$2\times 10^4$& $4\times 10^4$ &  $ 10^{-4}$ &$3.2\times 10^{-3}$&
$1.4\times 10^{-3}$& $1.4\times 10^{-7}$ &$4.4\times 10^{-11}$\\
$10^5$& $2\times 10^5$ &  $6\times 10^{-5}$ &$7.3\times 10^{-3}$& 
$3.2\times 10^{-3}$&$1.6\times 10^{-7}$& $8.3\times 10^{-11}$ \\
$10^5$& $2\times 10^5$ &  $ 10^{-4}$ &$5.6\times 10^{-3}$& 
$2.5\times 10^{-3}$&$9.5\times 10^{-8}$& $7.4\times 10^{-11}$ \\
$10^5$& $10^6$ &  $6\times 10^{-5}$ &$7.3\times 10^{-3}$&
$6.3\times 10^{-3}$& $1.6\times 10^{-7}$& $8.4\times 10^{-11}$\\
$10^5$& $10^6$ &  $10^{-4}$ &$5.6\times 10^{-3}$&
$4.8\times 10^{-3}$& $9.5\times 10^{-8}$&$7.4\times 10^{-11}$\\ \hline
\end{tabular}
\end{center}
{\footnotesize Table 1 ~The $CP$ asymmetry 
$\varepsilon$ and the baryon number asymmetry $Y_B(z_{EW})$ 
for several $M_{1,2}$ and $|\lambda_5|$. $M_N$ and $h_N$ are
fixed at $10^4$ GeV and $6\times 10^{-8}$, respectively.
The Yukawa coupling constants $h_{1,2}$ for $\Sigma_{1,2}$ are 
determined by the neutrino oscillation data. }
\end{figure}

\normalsize
Another interesting issue of the model is what is a lower bound of 
$M_N$ for which the required value of $Y_B(z_{EW})$ can be obtained. 
At $h_N=6\times 10^{-8}$ in the left panel of Fig.~5, $Y_B(z_{EW})$ is plotted 
by asterisks for $M_N=10^4$ GeV, 
$M_1=10^5$ GeV and $M_2=2\times 10^5$ GeV changing the value of 
$|\lambda_5|$ downward as $6\times 10^{-5}, 10^{-4}, 6\times 10^{-4}$.
In order to show what causes the difference among the cases 
with $M_N=10^4$ GeV and $h_N=6\times 10^{-8}$, 
we list parameters relevant to the leptogenesis in Table 1.
This suggests that the lower bound of  $M_N$ could be $10^4$ GeV
at least in the present model.\footnote{The possibility of low scale leptogenesis 
in the scotogenic type I seesaw has been intensively studied in \cite{kai}.
They concluded $M_N~{^>_\sim}~10^4$~GeV for the successful leptogenesis 
just assuming $N$ is in the thermal equilibrium initially.
Although we do not exhaust the parameter space, the similar bound 
of $M_N$ is obtained in the present model. } 
If the relevant parameters in the model are fixed at appropriate values which
can realize $|\varepsilon|~{^>_\sim}~10^{-7}$ and suppress the washout due to
$\Sigma_{\alpha}$ simultaneously at least for a sufficiently small $h_N$,
the low scale leptogenesis could be allowed in this model 
in a consistent way with the neutrino mass generation, 
the DM abundance and also the inflation.
We need no serious tuning for them even in that case.  

Finally, we remarks on the signatures in the collider experiment 
caused by the present low scale leptogenesis. 
Collider phenomenology expected for the triplet fermions has been 
discussed extensively in \cite{phenom}. 
Following it, any promising signature of the triplet fermions cannot be 
expected in the collider physics at least near future, 
since their masses should be larger than $O(10^4)$ GeV for the successful leptogenesis. On the other hand, even if the signatures of inert doublet 
scalars $\eta$ are discovered, it seems to be difficult to distinguish the scotogenic 
type III model from the scotogenic type I model.

\section{Summary}  
The scotogenic type III seesaw model is an interesting model
which can link the neutrino mass generation and the existence of DM.
Unfortunately, it cannot explain the baryon number asymmetry in the Universe
through the leptogenesis. Since heavy fermions in the model are triplets of 
$SU(2)$ and then have the gauge interaction, they are kept in the thermal 
equilibrium until the electroweak scale. As its consequence, they cannot generate 
the lepton number asymmetry through the out-of-equilibrium decay.    

We proposed a simple extension of the model by introducing a singlet fermion 
so as to incorporate successful leptogenesis.
Since this singlet fermion could be irrelevant to the neutrino mass generation
by assuming its Yukawa coupling constants are very small, its out-of-equilibrium 
decay could be possible even if it is not so heavy. 
If its thermal equilibrium could be prepared not through its Yukawa couplings 
but through other interaction, the leptogenesis caused by its decay at a
low temperature region could explain the required baryon number asymmetry.
As such a process, we supposed the singlet fermions pair production caused 
by the pair annihilation of  the triplet fermions which are produced 
in the inflaton decay.
Since the triplet fermions are in the thermal equilibrium at an early stage,
the singlet fermions could reach the thermal equilibrium at a high 
temperature where its equilibrium number density takes a large value.
Several parameter dependences of this leptogenesis were clarified in details.
We also showed that the required baryon number asymmetry could be generated
even for the small mass of the singlet fermion like $O(10^4)$~GeV as long as
the relevant parameters have suitable values. 
The scenario might be applicable for the low scale leptogenesis 
in other models for the neutrino mass, the DM and the inflation.

\section*{Acknowledgements}
This work is partially supported by MEXT Grant-in-Aid 
for Scientific Research on Innovative Areas (Grant No. 26104009)
and a Grant-in-Aid for Scientific Research (C) from Japan Society
for Promotion of Science (Grant No. 18K03644).

\newpage
\bibliographystyle{unsrt}

\end{document}